\documentclass[letterpaper, 10 pt, conference]{ieeeconf}  

\IEEEoverridecommandlockouts                              

\usepackage{hyperref} 
\usepackage{listings} 
\usepackage[backend=biber,style=ieee]{biblatex} 
\usepackage{booktabs}  
\usepackage{multirow} 
\usepackage{amsmath}
\usepackage{pgf} 
\usepackage{tikz} 
\usetikzlibrary{arrows.meta, positioning, fit, shapes.geometric, calc, backgrounds} 
\usepackage{makecell} 
\usepackage{amssymb}
\usepackage{footmisc}

\title{\LARGE \bf
Automating Information Extraction and Retrieval \\ for Industrial Spare Parts Pooling
}
\author{%
Dyuman Bulloni$^{1,}$\authorrefmark{2}, Rocco Felici$^{1,}$\authorrefmark{2}, Oliver Avram$^{1}$, and Anna Valente$^{1}$%
\thanks{$^{1}$Automation, Robotics and Machines (ARM), Institute of Systems and Technologies for Sustainable Production (ISTePS), University of Applied Sciences and Arts of Southern Switzerland (SUPSI), Lugano, Switzerland. {\tt \{dyuman.bulloni, rocco.felici, oliver.avram\}@supsi.ch}}%
\thanks{\authorrefmark{2} Equal contribution.}%
}

\begin{document}

\maketitle
\thispagestyle{empty}
\pagestyle{empty}

\begin{abstract}
Maintenance organizations in manufacturing try to avoid downtime and unnecessary purchasing by reusing existing assets, but the main obstacle is not a lack of parts but a lack of actionable visibility across sites and partners. Inventories are distributed, described with inconsistent naming conventions, and contain duplicates and partially specified references, so the right part often exists somewhere but remains effectively undiscoverable. 
The paper proposes PhRAG, a hybrid Retrieval-Augmented Generation for pooling this fragmented landscape into a Virtual Stock Pool (VSPool) that can be structured and searched as a single resource. Heterogeneous spare part descriptions are structured via Named Entity Recognition (NER) into a shared virtual pool dataset and indexed to support robust retrieval even when users express needs in natural language rather than exact technical specifications. The proposed modular pipeline leverages the multitasking nature of generative language models to cover two dimensions that make industrial parts pooling challenging: ($\boldsymbol{i}$) unstructured technical specifications from diverse data sources (e.g. new partners, catalogs, marketplace listings) are handled through an offline extraction and ($\boldsymbol{ii}$) request variability at runtime (references, partial references, specifications, price/condition constraints) is handled through a hybrid RAG-based search engine capable of retrieving relevant components and justifying results. The framework demonstrates the potential of generative approaches compared with traditional NER approaches in the presence of data scarcity for technical specifications extraction and overcomes the opacity of standard information retrieval systems  by generating justifications for retrieved components.
\end{abstract}
\section{INTRODUCTION}\label{sec:intro}
In the context of industrial maintenance, spare parts management aims to ensure the availability of components and reduce idle stock. Sharing spare parts inventories between plants or within a community could reduce the size of individual partners' stocks and establish a policy of dormant components circularity aimed at reducing overall waste. 
This objective introduces several technological challenges, in particular, ($i$) the need to create a common pool from several inventories, i.e., multiple unstructured data sources, and ($ii$) the ability to retrieve parts that meet the technical specifications of plants requiring maintenance.
The research presented in this publication aims to evaluate such challenges and proposes a framework to address them via a multitasking architecture relying on generative language models.

\setcounter{footnote}{1}
A virtual shared-pool inventory is required to accurately store relevant information, track quantities, and ensure that all stakeholders joining the community can easily access and order the exact parts needed. This gives rise to the need to structure a vast amount of textual data referenced with predefined entities from a highly technical domain via \textbf{technical specification extraction}. NER is one of the key techniques in Natural Language Processing (NLP) for information extraction tasks. Its main goal is to organize unstructured textual data into structured representations, improving the usability of such information. Although it is widely used in many contexts, the adaptation of such an approach for domain-specific applications is too often a repetitive process that involves continuous iterations of data labeling, training, and model evaluation~\cite{nadeau2007survey, lample2016neural}. In particular, in the manufacturing domain, the presence of highly specific terminology makes it difficult to train models capable of extracting information from unstructured textual data~\cite{Kumar2022FabNER}. 
Furthermore, solutions of this kind are heavily dependent on the creation of large amounts of annotated data~\cite{armingaud2025manufactubert}. In this research, we propose a language model-based approach that relies on large pre-trained models and their autoregressive generation of text to solve a NER task on data highly specific to the manufacturing domain. 
To ensure reproducibility, the paper describes the applied methodology, provides an open-source version of the implementation\footnote{\label{fn:code_note}Code and related material are publicly available at \href{https://github.com/automation-robotics-machines/PhRAG}{https://github.com/automation-robotics-machines/PhRAG}.}, and displays findings on FabNER~\cite{Kumar2022FabNER}, a publicly available benchmark dataset within the same domain. 

Regarding the \textbf{retrieval of relevant spare parts}, a search engine tailored for this application has been developed to map textual requests containing technical requirements of the user's needs to the data contained in the shared structured inventory.
It is implemented by modularly reusing components already developed for the extraction of technical specifications, demonstrating the ability of language models to adapt to different tasks within the same technical domain.
Its implementation focuses on efficiency and low-latency inference suitable for industrial deployment and interactive querying by the user. The evaluation of the retrieval capabilities is conducted on the proprietary dataset VSPool. 

The contributions of this work are twofold. Methodologically, we demonstrate that hybrid retrieval-augmented few-shot prompting substantially improves base models for NER in data-scarce technical domains, approaching supervised performance for low-resource entity types. At the system level, we present a modular pipeline whose shared components, i.e., embedding models, rank fusion, and generative model, serve both the offline extraction and the runtime search engine without requiring task-specific fine-tuning.

\paragraph*{Organization} 
We first discuss related work in Sec.~\ref{sec:rel_work}. Then present our framework in Sec.~\ref{sec:methods}, discussing how to solve the technical specifications extraction task and the components retrieval problem in a modular fashion. In Sec.~\ref {sec:results} we evaluate our method for the two tasks. Finally, we conclude by proposing future developments in Sec.~\ref{sec:conclusion}. 

\section{RELATED WORK}\label{sec:rel_work}

\subsection{Technical Specifications Extraction}

Classical NER approaches~\cite{Kumar2022FabNER} that tackle highly specialized domains are primarily supervised and rely on labeled corpora to learn mappings between textual data and semantic categories.

Early NER systems were predominantly rule-based and were used to identify entities with well-defined lexical or syntactic patterns, e.g., leveraging regular expressions~\cite{nadeau2007survey}. Such approaches are still widely used in industrial settings, particularly in constrained domains where entity formats are stable, and domain knowledge can be explicitly encoded. However, rule-based systems typically suffer from limited scalability and poor generalization to unseen variations.
To overcome such limitations, the proposed approach described in Sec.~\ref{sec:methods} relies on language models that, leveraging large-scale pre-training, provide generalist capabilities on NLP tasks and rapid adaptability to unseen linguistic problems.

With the availability of annotated data, neural approaches became dominant. Recurrent neural architectures such as Bidirectional Long Short-Term Memory networks (BiLSTMs), often combined with Conditional Random Fields (CRFs), demonstrated strong performance by modeling sequential dependencies and contextual information~\cite{lample2016neural}. 
Despite their success, supervised neural NER models remain highly dependent on annotated data and predefined entity schemas. 


More recently, transformer-based models have achieved state-of-the-art results in NER by leveraging deep contextualized word representations learned through large-scale pretraining.
The latest paradigm leverages Large Language Models (LLMs), which exhibit strong contextual reasoning capabilities and can handle unseen or loosely defined entity types. Unlike traditional NER models, LLM-based approaches are flexible with respect to data availability, supporting zero-shot learning, few-shot learning, and standard supervised learning. 
These models can perform entity extraction through prompting or lightweight fine-tuning, reducing the need for task-specific annotation~\cite{wang2025gpt,sainz2023gollie}. However, LLM-based approaches introduce new challenges, including a trade-off between response time and output quality, sensitivity to prompt design, and the occurrence of hallucination-generated outputs that are not grounded in the input data~\cite{wang2025slot}. 
To this end, the framework proposed includes adaptation to the manufacturing domain via a few-shot in-context examples compatible with both small and large language models. The latter could be used during the offline phase, and the former during the deployment of architectures designed for runtime and user interaction.

NER in manufacturing operates in a different regime from typical cases: texts are generally noisy, using technical domain knowledge as context for abbreviations, metrics, and a lack of cohesion for even the general categories. As a result of that, models trained on standard datasets perform poorly due to domain shift and a high rate of rare or unseen terms. \\
Entity types in this domain usually include components, materials, machines, parameters, and defects. Unlike traditional NER categories (e.g., PER, ORG, LOC), these entities are highly specialized and often appear as compound phrases, making token boundary detection more challenging. 
Research on the topic spans from fine-tuned architectures for guideline-following~\cite{sainz2023gollie} to continual pretraining on curated manufacturing data~\cite{armingaud2025manufactubert}.
To tackle such challenges, we opted for a hybrid RAG approach to gather in-context examples for a reliable NER in a specific domain. Unlike semantic-only retrieval, the hybrid approach is especially valuable in manufacturing because it preserves details such as reference excerpts and manufacturer codes, i.e. identifiers with no intrinsic semantic meaning that semantic embeddings typically fail to capture.

\subsection{Components Retrieval}
%
%
%
Language models are increasingly integrated into information retrieval pipelines within RAG systems. 
In such architectures, language models are often employed as rerankers to refine document ranking by assessing query–document relevance through their semantic understanding~\cite{gao2023chat}. 
Recent works explore more flexible and modular RAG frameworks, in which specialized components, such as retrievers, rerankers, and generators, are combined to improve retrieval quality and adaptability to domain-specific tasks~\cite{gao2024retrievalaugmentedgenerationlargelanguage}.
Building on this line of research, we propose a search engine adapted for the manufacturing domain that leverages a language model within a modular RAG to both rerank retrieved components and provide justifications for the results, thereby improving transparency.
The search engine is implemented to reuse, as shared modules, the embedding models, rank fusion method, and generative language model introduced in the \textit{technical specifications extraction} component above.


\section{PROPOSED APPROACH}\label{sec:methods}
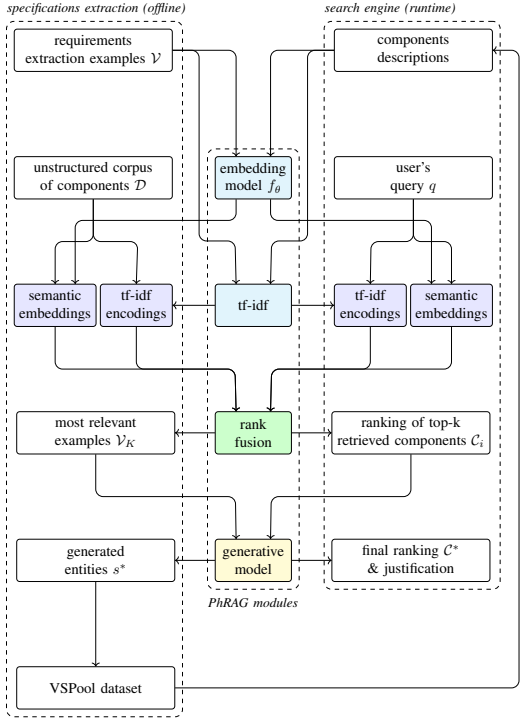
\begin{figure}[htbp]
\centering
\resizebox{.8\linewidth}{!}{
    \begin{tikzpicture}[
      box/.style={rectangle, draw, rounded corners=2pt, minimum width=1.8cm, minimum height=1cm, text centered, align=center},
      group/.style={draw, dashed, inner sep=5pt, rounded corners=5pt},
      node distance=2cm and 1cm
    ]

    \node[box, text width=3.5cm] (examples) {requirements extraction examples $\mathcal{V}$};
    \node[box, text width=3.5cm, below=of examples] (corpus) {unstructured corpus of components $\mathcal{D}$};
    \node[box, fill=blue!10, minimum width=1.7cm, below=2cm of corpus.south west, anchor=north west] (edb) {semantic\\embeddings};
    \node[box, fill=blue!10, minimum width=1.7cm, below=2cm of corpus.south east, anchor=north east] (indexed) {tf-idf\\encodings};
    \node[box, text width=3.5cm, below=2cm of $(edb.south)!0.5!(indexed.south)$, anchor=north] (mre) {most relevant\\examples $\mathcal{V}_K$};
    \node[box, text width=3.5cm,  below=of mre] (ge) {generated\\entities $s^*$};

    \node[right=of examples,minimum width=4cm] (sec_col) {}; 
    \node[box, fill=cyan!10, right=of corpus] (em) {embedding\\model $f_{\theta}$};
    \node[box, fill=cyan!10, below=of em] (tfidf) {tf-idf};

    \node[box, fill=green!20, below=of tfidf] (rrf) {rank\\fusion};
    \node[box, fill=yellow!20, below=of rrf] (gm) {generative\\model};

    \node[box, text width=3.5cm, below=of ge] (vspool) {VSPool dataset};

    \node[box, right=of em, text width=3.5cm] (query) {user's\\query $q$}; 
    \node[box, above=of query, text width=3.5cm] (components) {components\\descriptions}; 
    \node[box, fill=blue!10, minimum width=1.7cm, below=2cm of query.south west, anchor=north west] (indexed1) {tf-idf\\encodings};
    \node[box, fill=blue!10, minimum width=1.7cm, below=2cm of query.south east, anchor=north east] (edb1) {semantic\\embeddings};
    \node[box, text width=3.5cm, below=2cm of $(edb1.south)!0.5!(indexed1.south)$, anchor=north] (test1) {ranking of top-k retrieved components $\mathcal{C}_i$};
    \node[box,text width=3.5cm, below=of test1] (test2) {final ranking $\mathcal{C}^*$\\\& justification};

    \draw[->, rounded corners=5pt] (examples) -| ($(em.north west) + (.5cm,0)$);
    \draw[->, rounded corners=5pt](examples) -| ($(examples.south)+(2.5,-4)$) -- ($(examples.south)+(3.35,-4)$) -- ($(tfidf.north west) + (.5cm,0)$);
    \draw[->, rounded corners=5pt] (components) -| ($(em.north east) - (.5cm,0)$);
    \draw[->, rounded corners=5pt](components) -| ($(components.south)-(2.5,4)$) -- ($(components.south)-(3.35,4)$) -- ($(tfidf.north east) - (.5cm,0)$);

    \draw[->, rounded corners=5pt](corpus) -- ($(corpus.south)-(0,1)$) -- ($(corpus.south)-(.9,1)$) -- (edb.north);
    \draw[->, rounded corners=5pt](corpus) -- ($(corpus.south)-(0,1)$) -- ($(corpus.south)+(1.05,-1)$) -- (indexed.north);

    \draw[->, rounded corners=5pt](query) -- ($(query.south)-(0,1)$) -- ($(query.south)-(1,1)$) -- (indexed1.north);
    \draw[->, rounded corners=5pt](query) -- ($(query.south)-(0,1)$) -- ($(query.south)+(.9,-1)$) -- (edb1.north);
    
    \draw[->] (tfidf)  to[out=180,in=0]  (indexed);
    \draw[->] (tfidf)  to[out=0,in=180]  (indexed1);

    \draw[->, rounded corners=5pt] ($(em.south west) + (.5cm,0)$) -- ($(em.south west) + (.5cm,-.5)$) -- ($(edb.north east)+(-.5cm,1.5)$) --  ($(edb.north east) - (.5cm,0)$);
    \draw[->, rounded corners=5pt] ($(em.south east) - (.5cm,0)$) -- ($(em.south east) - (.5cm,.5)$) -- ($(edb1.north west)+(.5cm,1.5)$) --  ($(edb1.north west) + (.5cm,0)$);

    \draw[->, rounded corners=5pt](indexed) -- ($(indexed.south)-(0,1)$) -- ($(edb.south)+(4.25,-1)$) -- ($(rrf.north west) + (.5cm,0)$);
    \draw[->, rounded corners=5pt](edb) -- ($(edb.south)-(0,1)$) -- ($(edb.south)+(4.25,-1)$) -- ($(rrf.north west) + (.5cm,0)$);
    
    \draw[->, rounded corners=5pt](indexed1) -- ($(indexed1.south)-(0,1)$) -- ($(edb1.south)+(-4.25,-1)$) -- ($(rrf.north east) - (.5cm,0)$);
    \draw[->, rounded corners=5pt](edb1) -- ($(edb1.south)-(0,1)$) -- ($(edb1.south)+(-4.25,-1)$) -- ($(rrf.north east) - (.5cm,0)$);

    \draw[->] (rrf) -- (mre);
    \draw[->] (rrf) -- (test1);
    \draw[->, rounded corners=5pt](mre) -- ($(mre.south)-(0,1)$) -- ($(mre.south)-(-3.3,1)$) -- ($(gm.north west) + (.5cm,0)$);
    \draw[->, rounded corners=5pt](test1) -- ($(test1.south)-(0,1)$) -- ($(test1.south)+(-3.3,-1)$) -- ($(gm.north east) - (.5cm,0)$);
    \draw[->] (gm) -- (ge);
    \draw[->] (ge) -- (vspool);
    \draw[->, rounded corners=5pt](vspool) -- ($(vspool)+(10cm,0)$) |- (components);
    \draw[->] (gm) -- (test2);
    
    \node[group, fit=(examples)(indexed)(ge)(edb)(vspool)] (offline) {};
    \node at ($(offline.north west)+(-0.1,0.3)$) [anchor=west] {\small \textit{specifications extraction (offline)}};
    \node[group, fit=(em)(rrf)(gm)(tfidf)] (modules) {};
    \node at ($(modules.south west)+(-0.1,-0.3)$) [anchor=west] {\small \textit{PhRAG modules}};
    \node[group, fit=(components)(test2)] (runtime) {};
    \node at ($(runtime.north west)+(-0.1,0.3)$) [anchor=west] {\small \textit{search engine (runtime)}};

    \end{tikzpicture}
}

\caption{PhRAG modular framework. Specifications extraction on the left (Sec.~\ref{subsec:tec_spe_extraction}), search engine on the right (Sec.~\ref{subsec:sea_method}) and shared components in the central column.}
\label{fig:architecture}
\end{figure}

The PhRAG framework proposed consists of two phases, the \emph{technical specifications extraction} (Sec.~\ref{subsec:tec_spe_extraction}) and the \emph{search engine} (Sec.~\ref{subsec:sea_method}). The former, as illustrated in the left branch of Fig.~\ref{fig:architecture}, takes place offline and integrates unstructured data $\mathcal{D}$ from various sources via NER into the Virtual Stock Pool (VSPool) dataset. The engine instead (right column in Fig.~\ref{fig:architecture}) queries the VSPool for the retrieval of spare parts whenever a user, looking for a component, interacts with the system. The two modules, while remaining decoupled at runtime, share the use of pre-trained components and exclude fine-tuning, proposing a few-shot approach whose examples are identified and employed through a hybrid RAG~\cite{lewis2020retrieval}. 

\subsection{Technical specifications extraction}\label{subsec:tec_spe_extraction}
%
One of the fundamental requirements of the proposed framework is the definition of a common, consistent data representation that remains compatible across heterogeneous sources, namely the VSPool dataset. To this end, the technical specifications extraction is structured to enable the systematic collection and organization of technical data.

Formally, we are given a corpus $\mathcal{D}=\{d_i\}_{i=1}^N$ of unstructured textual descriptions of industrial components originating from different sources, and a finite, pre-defined collection $\mathcal{S}=\{s_j\}_{j=1}^M$ of technical specification types of interest. The task of technical specification extraction can be formulated as a NER problem.
Each description $d\in \mathcal{D}$ is represented as a sequence of tokens $d=(t_1,t_2, \dots, t_{|d|})$. The goal is to assign to each token $t_k$ a label $s_k$ from a label set derived from $\mathcal{S}$, producing a corresponding sequence of labels $s^*=(s_1,s_2,\dots,s_{|d|})$, referred as generated entities in the diagram above.

For the annotation of each description $d$ in the corpus, we rely on a RAG.
We retrieve a set $\mathcal{V}_K=\{ v_k\}_{k=1}^K\subseteq \mathcal{V}$ of $K$ examples from a cache $\mathcal{V}$ of pre-annotated descriptions. 
Such examples comprise pairs of tokens and correspondent technical specifications, $v = \{(t_i,s_i)\}_{i=1}^{|d|}, t_i \in d,\; s_i \in \mathcal{S}$. 

The retrieval is defined as the pipeline $\phi(\mathcal{V}_K|d)$ that, given the description $d$ of a component, provides the set of examples that are most relevant to that description.
The retrieval provides contextual information that, through few-shot prompting~\cite{brown2020language}, guides the generation towards more reliable extraction.

The generation returns the sequence of technical requirements extracted $s^*=(s_1,s_2,\dots,s_{|d|})$ associated to the tokenized description $d=(t_1,t_2,\dots,t_{|d|})$ as follows:
\begin{equation}
    p_{\eta}(s^*|d,\mathcal{V}_K)=\prod_{i=1}^{|d|}p_{\eta}(s_i|d,\mathcal{V}_K,s_{i-1}).
\end{equation}

\subsubsection*{Examples retrieval}
Retrieval, in the context of the NER problem, refers to selecting the most relevant examples that can guide, via few-shot prompting, the language model in generating the appropriate sequence of labels.
In the proposed approach, retrieval is defined in a hybrid manner that combines semantic and lexical retrievers. Given a set of annotated descriptions (referred as $\mathcal{V}$ in Fig.~\ref{fig:architecture}), these are used to create tf-idf sparse encodings and semantic dense encodings. 
Semantic embeddings are created $x_i=f_{\theta}(v_i)\in \mathbb{R}^n, v_i \in \mathcal{V}$ using a pretrained transformer-based embedding model $f_\theta$ with parameters $\theta$. 
The embedding model $f_\theta$ used is  all-MiniLM-L6-v2~\cite{wang2020minilmdeepselfattentiondistillation}, an uncased sentence embedding model distilled from BERT with $12$ layers, hidden size of $384$, and $33$M parameters. 

For the retrieval of in-context examples $\mathcal{V}_K$, the description $d$ is embedded and compared using the Okapi Best Matching (BM25) function~\cite{robertson2009probabilistic} to provide a relevance score to find more relevant examples via sparse embeddings.
$\mathcal{V}_{K}^\text{BM25}$ is the set of $K$ annotated descriptions with the highest BM25 scores.
A second set of relevant examples $\mathcal{V}_{K}^\text{SEM}$ is obtained by comparing the dense embedding of the description $d$ against embeddings of component descriptions in $\mathcal{V}$. Comparison between embeddings is done via $k$-nearest neighbors and similarity score between description embeddings $x,y$, L2-normalized as $\hat{x}=\frac{x}{\lVert x \rVert_2},\hat{y}=\frac{y}{\lVert y \rVert_2}$, is defined as cosine similarity: $S_{\text{cos}}(x,y) = \hat{x}^\top\hat{y}$. 
Both sets of annotated descriptions $\mathcal{V}_{K}^\text{BM25}$ and $\mathcal{V}_{K}^\text{SEM}$ are ranked from most to least relevant.
%
%
Defined $\mathcal{R}=\{r^\text{BM25}, r^\text{SEM}\}$ as the set of the rank functions of both $\mathcal{V}_{K}^\text{BM25}$ and $\mathcal{V}_{K}^\text{SEM}$, to find the optimal intersection $\mathcal{V}_K$ between the two, Reciprocal Rank Fusion
(RRF)~\cite{cormack2009reciprocal} has been employed: 
$$
\mathrm{RRF}\big(v \in \mathcal{V}_K^{\text{BM25}} \cup \mathcal{V}_K^{\text{SEM}}\big) = \sum_{r \in R} \frac{1}{b + r(v)}
$$
where $b=60$.
The ranking provided by RRF scores defines the examples $\mathcal{V}_K$.

\subsubsection*{Entities generation}
Examples in $\mathcal{V}_K$ comprise sequences of token-technical specification pairs $(t,s)$ for all tokens of each description.
A decoder-only language model~\cite{roziere2023code} is prompted with information about the NER task to be solved, the definition of the entities to be generated, the $K$ examples and $d=(t_1,t_2,\dots,t_{|d|})$, and the unseen description to annotate. The generation returns the sequence of technical specifications extracted $s^*=(s_1,s_2,\dots,s_{|d|})$ associated to each token of $d$. At each step, the generation is constrained to available technical specification $\mathcal{S}$, masking generations of invalid specifications with $\text{logit}=-\infty$. Thus guaranteeing that the generated sequence remains valid

\subsection{Search engine}\label{subsec:sea_method}
As presented in the previous Sec.~\ref{subsec:tec_spe_extraction}, technical specifications extraction is crucial to join together unstructured technical data from different plants. 
Once the categories and technical specifications of interest are defined as $\mathcal{S}=\{s_j\}_{j=1}^M$, we construct the VSPool, a dataset of industrial components $\mathcal{C}=\{c_i\}_{i=1}^N$. Each component $c_i$ is associated with a set of technical specifications $\mathcal{S}_i \subseteq \mathcal{S}$, yielding the dataset representation, $
    \text{VSPool}=\{(c_i,\mathcal{S}_i)\}_{i=1}^N
$.
The pool of components is shared among all partners, and the search engine draws from such a large collection of structured information regarding available components across the whole community.
The engine is implemented to allow the retrieval of spare parts that a certain partner may need. 
Allowing any partner to find components among the shared inventory is necessary to ensure circularity of dormant stock across the part-sharing community. 

In practice, the search engine accepts as input a query $q$ of the user in natural language, against which it provides a ranked list of relevant spare parts $\mathcal{C}_i \subseteq\mathcal{C}$ that meet the technical needs of the user. The engine is based on a search service and a graphical interface that includes a search bar, filters, and facets. 
The objective is, in fact, an information retrieval task and was addressed through the use of a RAG.

The retrieval consists of a pipeline $\phi(\mathcal{C}_i|q)$ that, given the query $q$, provides a preliminary ranked list $\mathcal{C}_i$. 
Unlike, the generation for the technical specifications extraction (Sec.~\ref{subsec:tec_spe_extraction}), here no example is given instead, in a zero-shot manner, the ranked list is prepended to the prompt of the language model that is employed as reranker to provide an optimal final ranking $\mathcal{C}^*$ and transparency in the form of an explicit justification of the provided ranking.

\subsubsection*{Retrieval}
Given VSPool, in the form that will be detailed in Sec.~\ref{subsec:datasets}, the retrieval for the search engine focuses on the following variables: \emph{reference}, \emph{name} and \emph{description} of components.
The retrieval consists of a hybrid approach, i.e., during an offline phase, these data are encoded via tf-idf and embedded in dense vectors of dimension $384$ using the pretrained model all-MiniLM-L6-v2.
Data is then indexed in an Elasticsearch cluster~\cite{elasticsearch}.

At runtime, the lexical search is focused on references and keywords. The implementation is based on the BM25 function to provide a relevance score function for the ranking of the components retrieved. 
Given a query $q$ from the user, a spare part name, reference, or description $d$ from the set of dormant stock VSPool, the BM25 score for such components is defined as:
\begin{equation*}
    \text{BM25}(d)=\sum_{i\in q} \log \frac{N}{df_i}\times\frac{(k_1+1)tf_i}{k_1((1-b)+b\frac{dl}{avdl})+tf_i}
\end{equation*}
where $\log \frac{N}{df_i}$ is the idf, i.e., the ratio between the total number of spare parts in the whole inventory of dormant stock and the number of spare parts descriptions that contain the $i$-th term of the query $q$;
$tf_i$ defines the frequency of the $i$-th term of query $q$ in the spare part description $d$;
$dl = \sum_{i\in d} tf_i$ is the length of description $d$ of a component, $avdl$, the average length	of spare parts descriptions over the whole inventory.
Parameters $b\in[0,1]$ and $k\in\mathbb{R}_{\geq0}$ are set to $0{.}75$ and $1.5$ respectively. 
BM25, with respect to other traditional approaches based on tf-idf encodings, integrates document length into the scoring but essentially still relies heavily on exact word matching. 
To overcome this limitation, the Levenshtein edit distance with parameter $l=2$ is introduced, allowing up to two editing operations (namely substitution, insertion, or deletion of a character) during word comparison.
%

Reference, name, and description of spare parts $d_i\in C$ from the inventory are also embedded in a latent embedding, i.e., a dense vectorial representation, defined as $x_i=f_{\theta}(d_i)\in \mathbb{R}^d$ using a pretrained language embedding model $f_\theta$ with parameters $\theta$. The similarity between a query $q$ of a user interacting with the search engine and a description $d_i$ of the $i$-th component is measured based on the corresponding embeddings $x,y$ L2-normalized as $\hat{x}=\frac{x}{\lVert x \rVert_2},\hat{y}=\frac{y}{\lVert y \rVert_2}$, is defined as cosine similarity:$S_{\text{cos}}(x,y) = \hat{x}^\top\hat{y}$.
The comparison of embeddings is done via $10$-nearest neighbors using cosine similarity as a distance
measure between vectors.


Both retrieval approaches output a ranking of top-$n$ relevant spare parts $\mathcal{C}_n^{\text{LEX}},\mathcal{C}_n^{\text{SEM}}$; a combination of rankings from the two is implemented defining RRF as: 
\begin{equation}\label{eq:rrf_search}
    \mathrm{RRF}\big(c \in \mathcal{C}_n^{\text{LEX}} \cup \mathcal{C}_n^{\text{SEM}} \big) = \sum_{r_i \in R}^{N} \frac{w_i}{b + r_i(c)}
\end{equation}
where $R=\{r^{\text{LEX}}, r^{\text{SEM}}\}$ is the set of ranking functions, $w_i=1, \forall i$ to ensure equal contribution of the lexical-based ranking and the semantic one, and the parameter $b$ is set to $60$.

\subsubsection*{Filtering}
At run-time, the users can either interact with the system through a
search bar providing a free text query $q$ in natural language or selects filters and facets to explicitly define preferences regarding the spare parts they are looking for. 
Filtering mechanisms have been implemented to refine the ranking output by the search modules $\phi(\mathcal{C}_i|q,f)$ only to the subset of components that would meet the requisites $f$ of the user. Filters $f$ are 
manually applied by the user or automatically applied whenever $q$ contains technical information. Filters on VSPool are available for the following variables: \emph{condition}, \emph{brand}, \emph{categories}, and \emph{price} range.
Automatic filtering is implemented with low-latency entities extractions, fine-tuning the en\_core\_web\_sm~\cite{spacy2} pipeline on ad-hoc datasets~\cite{snapnets} ($1{.}371$ annotated queries for conditions, $100$ annotated instances for prices, $373$ queries for brand and categories). 
\subsubsection*{Reranker}
The set of ranked pertinent spare parts retrieved is then refined by a language model prompted in a zero-shot manner. A pre-trained model, Llama $3.1$~\cite{grattafiori2024llama}, with $7$B parameters running locally is used. The prompt is formed by the query of the user $q$, the ranking retrieved $\mathcal{C}_i$, and a detailed description of the task. The generation includes the proposal of a definitive ranking, $\mathcal{C}^*$, and a detailed justification for the choice. 
Compared to traditional information retrieval systems, the proposed approach retrieves relevant results, providing technical justifications of the choice to the user, discussing their requirements against the components retrieved.
Not only is the explanation useful to provide transparency of the system
for the user, but it also improves the performance of the model in ranking the best parts first. 


\section{RESULTS}\label{sec:results}
\subsection{Datasets}\label{subsec:datasets}
For the evaluation of the technical specifications extraction introduced in Sec.~\ref{subsec:tec_spe_extraction}, it is considered FabNER~\cite{Kumar2022FabNER}, which is a collection of datasets of manufacturing text comprising more than $350{.}000$ words for NER purposes. Unstructured text is annotated for every word with entities representing categories of industrial processes: Material (MATE), Manufacturing Process (MANP), Machine/Equipment (MACEQ), Application (APPL), Features (FEAT), Mechanical Properties (PRO), Characterization (CHAR), Parameters (PARA), Enabling Technology (ENAT), Concept/Principles (CONPRI), Manufacturing Standards (MANS), and BioMedical (BIOP).
In particular, for the validation, we referred to the FabNER-simple dataset composed of $9{.}435$ training instances and $2{.}064$ test ones. The dataset consists of an id for each sentence (string), the list of tokens of the sentence (list of strings), and the list of entity tags (list of integers).%

The Virtual Stock Pool (VSPool) dataset has been constructed for validation purposes of the search engine. The dataset comprises $4{.}020$ components for industrial processes across eleven brands. Spare parts were selected among the most representative brands on the market, i.e., among the top eleven brands in terms of the number of components available on component providers' marketplaces. Selection has been made to provide variability in terms of industrial processes, reference systems across several distributors and constructors. Each component presents the following variables: query (free text), reference (alphanumeric id, unique), categories (list of strings, keywords), brand (string, keyword), condition (string, keyword), general discursive description (free text), description embedding (dense vector), name or short description (free text), name embedding (dense vector), price (float, currency in euros), technical specifications (key, field, unit, value) respectively in the format (string, string, string, float), set of pictures, attachments. The queries were collected to represent the range of natural language expressions an operator might use to look for a component. Queries were collected via a questionnaire\footref{fn:code_note} distributed to industry experts and professionals and then synthetically augmented.
Categories and facets are defined a priori based on hierarchical classification schemes. 
The degree of specialization increases as category subsets become progressively narrower. For example, the category \emph{electrical equipment} includes the subcategory \emph{switches}, which in turn contains the more specialized subset \emph{toggle switches $\&$ accessories}.
Condition is a fixed string corresponding to different levels of item conditions, e.g., \emph{refurbished}, \emph{new}, \emph{used}. The description usually comprises the possible application of the spare part, the
general context of its usage, and some details on its characteristics. The price in euros is stored for each part without considering any taxation. Technical specifications are saved as a tuple of key, value pairs which include a description of the
specification, e.g., \emph{maximum flow} and the corresponding value and UoM, e.g., \emph{1.8 liters per second}.
Attachments comprise links to valid offers on marketplaces.

\subsection{Metrics}
The evaluation of the technical specification extraction is based on the FabNER-simple~\cite{Kumar2022FabNER} test set with $2{.}064$ annotated instances and relies on the following metrics:
\begin{equation*}
\begin{split}
    \text{Precision} &= \frac{TP}{TP+FP}, \quad \text{Recall} = \frac{TP}{TP+FN}, \\
    \text{F}_1\text{ score} &= 2\times\frac{\text{Precision}\times\text{Recall}}{\text{Precision}+\text{Recall}},
    \end{split}
\end{equation*}
where true positive ($TP$) refers to the entities correctly predicted, false positive ($FP$) to an entity predicted but not present in the ground truth, and false negative ($FN$) to an entity not predicted but present in the ground truth.%

Instead, the assessment of the search engine is based on the VSPool dataset (described in Sec.~\ref{subsec:datasets}) and consists of the evaluation of the ranked retrieval results that a search provides. In a supervised setting, the accuracy at k (or A@k) is defined as the number of times where the correct components are found among the top k results in the predicted ranking. Although not exhaustive, leaving room for future work to further strengthen the evaluation with other metrics, A@k sufficiently reflects the user expectation of retrieving relevant components at the top of the ranking.

\subsection{Evaluation} 

The evaluation is divided into two phases: the test of the technical specification extraction  (Sec.~\ref{subsec:ner_eval}) on the FabNER test set and the analysis of the performance of the search engine (Sec.~\ref{subsec:sea_eval}) on the VSPool dataset. Experiments were conducted on a Lenovo 82WQ with a dedicated NVIDIA GeForce RTX 4080 Laptop GPU running Ubuntu 22.04.4 LTS.

\subsubsection{Technical specification extraction evaluation}\label{subsec:ner_eval}

The first experiment involves comparing generative language-based models with their PhRAG version to quantify the added value of the framework detailed in Sec.~\ref{subsec:tec_spe_extraction}. 
Small Language Models (SLMs)~\cite{roziere2023code} and an LLM\footnote{We consider small models with less than $10$B parameters and large models with more than $100$B parameters.} are selected.
Base models are prompted with the same task definition, entities definition, and static few-shot examples. 
The same models are employed in the corresponding PhRAG.
Temperature is set to zero for all models to provide reproducibility of the results.
Up to five iterations with different pseudo-random seeds are executed to evaluate variability.
An absent standard deviation indicates negligible variability between runs.

\begin{table}[ht]
\centering
\resizebox{\linewidth}{!}{
\begin{tabular}{lcccccc}
   \toprule
    & \multicolumn{4}{c}{SLM}&\multicolumn{2}{c}{LLM}\\
    \cmidrule(lr{4pt}){2-5}
    \cmidrule(lr{4pt}){6-7}
        & Phi-$3{.}5$-mini
        & PhRAG $3{.}5$B 
        & \makecell{Code\\Llama $7$B} 
        & PhRAG $7$B 
        & GPT-5.2 
        & PhRAG \\
    \midrule
    \makecell{$\text{F}_1\text{ score}$\\avg $\pm$ std}  
        & $0.04$ 
        & \underline{$0.10$} 
        & $0.06\pm0.01$
        &  \underline{$0.19\pm0.2$}
        & $0.36\pm0.016$  
        & \underline{$0.59\pm0.012$}\\
    \bottomrule
\end{tabular}
} 
\caption{Comparison between base models and PhRAG \\ for $3.5$B, $7$B, and $100$B$+$ parameters on FabNER.}
\label{tab:base_proposed}
\end{table}
Results are presented in Tab.~\ref{tab:base_proposed} in terms of weighted average F1 scores across all entities in the FabNER test set. The findings indicate performance improvements across language models of all sizes. The improvements are more pronounced for larger models. Notably, when PhRAG is integrated with a state-of-the-art LLM, a substantial performance gain is observed, achieving an improvement of up to $23$ percentage points over the base model.



%
A second analysis focuses on evaluating the gap that persists between few-shot prompted generalist models and strictly supervised approaches that, fine-tuned on the task, represent the state of the art. 
The evaluation is important because, although supervised approaches generally achieve superior performance, their effectiveness is constrained by the availability of training data, which is not always accessible in the manufacturing industry.
Supervised approaches are trained on $9.44$k documents annotated with technical specifications.
Different approaches have been evaluated on the FabNER~\cite{Kumar2022FabNER} test set described in Sec.~\ref{subsec:datasets}.
Selected approaches for the validation are: small pretrained generative language models or SLMs (Phi-3.5-mini, Code Llama $7$B~\cite{roziere2023code}), large language models (a fine-tuned model GoLLIE $34$B~\cite{sainz2023gollie} for guidelines following and GPT-5.2-2025-12-11), and an ad-hoc trained pipeline implemented or replicated from state of the art to serve as a benchmark: SpaCy en\_core\_web\_sm pipeline and BiLSTM+CRF~\cite{Kumar2022FabNER}.
\begin{table}[ht]
\centering
\resizebox{\linewidth}{!}{
\begin{tabular}{lccccccc}
   \toprule
    & \multicolumn{2}{c}{SLM}&\multicolumn{3}{c}{LLM}&\multicolumn{2}{c}{Supervised training}\\
    \cmidrule(lr{4pt}){2-3}
    \cmidrule(lr{4pt}){4-6}
    \cmidrule(lr{4pt}){7-8}
        & Phi-$3{.}5$-mini
        & \makecell{Code\\Llama $7$B} 
        & GoLLIE $34$B
        & GPT-5.2 
        & PhRAG
        & SpaCy 
        & \makecell{BiLSTM\\+CRF}\\
    \midrule
    \makecell{$\text{F}_1\text{ score}$\\avg $\pm$ std}   & $0.04$ &    $0.06$ & $26.3\pm0.4$ & $0.36\pm0.016$  & $0.59$      &  $0.82 \pm 0.01$ & \underline{$0.88$}\\
    \bottomrule
\end{tabular}
}
\caption{Comparison between supervised models, generative approaches, and PhRAG on FabNER.}
\label{tab:bench}
\end{table}
Results in Tab.~\ref{tab:bench} provide the overall weighted average F1 score for the approaches in comparison. Scores are weighted with respect to the unbalanced classes of the entities. Temperature and number of iterations are kept as in the first experiment, respectively, zero and five.
Despite the enormous generalization capabilities of language models, the state of the art in technical specification extraction for the manufacturing domain remains predominantly dominated by approaches trained ad hoc for the task in question.
PhRAG, in its biggest size, significantly reduces the gap between generative and supervised approaches despite operating without task-specific fine-tuning. While supervised models still achieve the best performance, PhRAG shows that a modular retrieval-augmented design can substantially improve base models, LLM-based NER while also supporting the needs of the search engine described in Sec.~\ref{subsec:sea_method}.

\begin{figure}[ht]
\centering
\resizebox{\linewidth}{!}{
\input{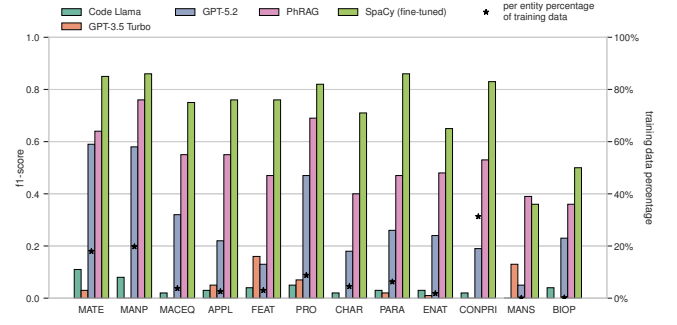}
}
\caption{Entity-level comparison between supervised approaches, generative approaches, and PhRAG on FabNER.}
\label{fig:ie_comp}
\end{figure}
In Fig.~\ref{fig:ie_comp}, are shown the results of Tab.~\ref{tab:bench}, in more detail, with an overview of performances across all entities comprised in the test set.
The plot reveals that the performance gap between PhRAG and supervised methods is not uniform across entity types.
For entities with high volume of training data, it is possible to notice that few-shot prompting approaches for SLMs still suffer from a large performance gap compared to ad-hoc trained technologies such as SpaCy for the task of technical information extraction: Code Llama $7$B presents, in the best case, a gap of $36$ percentage points in performance for the entity MANS and $83$-points difference for the entity PARA, which is the worst case scenario.
LLMs reduce the performance gap relative to smaller models and the state of the art.
Nonetheless, PhRAG furthermore narrows the difference between LLMs and supervised approaches and, in some cases, matches the benchmark supervised approaches.
For entities with limited training support, PhRAG matches or surpasses SpaCy, suggesting that retrieval-augmented few-shot prompting is most advantageous precisely where annotation budgets are lowest.
Data scarcity significantly affects the performance of supervised learning approaches. SpaCy's performances go below $0.60$ only for entities with scarce training data: MANS, $45$ annotated entities, and BIOP $107$.
In these extreme cases, e.g., MANS entity, PhRAG presents an improvement to the state of the art.
In a deployment scenario, this suggests a targeted annotation strategy: prioritizing labeling for the highest-impact entity types while relying on PhRAG for long-tail entities where supervision is infeasible.
Up to now, we have discussed only versions of PhRAG for base models, since in the manufacturing sector, curated data is not always available for fine-tuning. As a final remark, we underline that also fine-tuned language models can be employed in PhRAG. In this regard, a fine-tuned version (on FabNER training set) of Code Llama-$7$B model is obtained via Low-Rank Adaptation (LoRA)~\cite{DBLP:journals/corr/abs-2106-09685} and equipped with PhRAG. 
A drastic improvement is observed with respect to the base model (Tab.~\ref{tab:bench}), by more than $10$ times the original F1 score, reaching a $0.67$ weighted F1-score. An approach that combines PhRAG and an efficient fine-tuning on scarce data is left for future developments.

These results suggest a practical deployment guideline: when labeled data are available in sufficient quantity, supervised fine-tuning remains the preferred approach. PhRAG is most advantageous in two complementary scenarios i.e. the initial deployment with no task-specific annotations, and integration of new inventory sources where labeling a representative sample is impractical in the short term. As annotations accumulate, the LoRA-based variant of PhRAG would offer a clear upgrade path without requiring a full re-architecture.

\subsubsection{Search engine evaluation}\label{subsec:sea_eval}
The search engine introduced in Sec.~\ref{subsec:sea_method} has been evaluated in two respects. The first is the measure of the impact on performances of different embedding models; the second is an ablation study to determine the contribution of the different parts of the system. 
%
%
%
\paragraph*{Embedding models selection}
Key aspects in the choice of a suitable embedding model for the semantic search, presented in Sec.~\ref{subsec:sea_method}, are the dimension of the embeddings and the impact they have on the quality of semantic search. 
Larger embeddings have greater representational capacity but suffer from longer inference times for embedding creation. 
To balance these aspects, the impact of embedding models in the semantic search is evaluated in a supervised setting on the VSPool dataset described in Sec.~\ref{subsec:datasets}. 
The dataset comprises queries in natural language and corresponding components. The queries are embedded and employed in the semantic search to observe within how many positions $k$ in the ranking the spare part corresponding to the query would have been retrieved. Such an evaluation has been done for $k=1,5,10$ and averaging the results across these three values of $k$. Results are available in the Tab.~\ref{tab:embedding_model_evaluation}. The set of models evaluated represents the current state of the art~\cite{enevoldsen2025mmtebmassivemultilingualtext}.
\begin{table}[htbp]
\centering
\resizebox{\linewidth}{!}{
\begin{tabular}{lccccc}
    \toprule
    Model & \makecell{Embedding\\dimension} & A@$1$ & A@$5$ & A@$10$ &  avg A@$k$ \\
    \midrule
    bge-m3 & $1024$ & \underline{$0.716$} & \underline{$0.836$} & \underline{$0.875$} & \underline{$0.809$} \\
    all-MiniLM-L12-v2 & \underline{$384$} & $0.633$ & $0.798$ & $0.841$ & $0.757$ \\
    all-MiniLM-L6-v2 & \underline{$384$} & $0.627$ & $0.8$ & $0.842$ & $0.756$ \\
    gte-modernbert-base & $768$ & $0.627$ & $0.795$ & $0.84$ & $0.754$ \\
    gtr-t5-large & $768$ & $0.621$ & $0.796$ & $0.836$ & $0.751$ \\
    m3e-base & $768$ & $0.625$ & $0.789$ & $0.839$ & $0.751$ \\
    all-mpnet-base-v1 & $768$ & $0.6$ & $0.772$ & $0.836$ & $0.736$ \\
    multi-qa-mpnet-base-cos-v1 & $768$ & $0.593$ & $0.765$ & $0.822$ & $0.727$ \\
    all-mpnet-base-v2 & $768$ & $0.554$ & $0.734$ & $0.804$ & $0.697$ \\
    paraphrase-MiniLM-L6-v2 & \underline{$384$} & $0.548$ & $0.719$ & $0.778$ & $0.682$ \\
    \bottomrule
\end{tabular}
}
\caption{Embedding models evaluation:\\ spare parts found within top-$k$ results ($k=1,5,10$)}
\label{tab:embedding_model_evaluation}
\end{table}
Best results (underlined in the Tab.~\ref{tab:embedding_model_evaluation}) correspond to a model that has embeddings of size $1024$, which slows down the process for the creation of the vectors. The fastest model is all-MiniLM-L6-v2~\cite{wang2020minilmdeepselfattentiondistillation} that still presents optimal performance in terms of parts found when employed in the semantic search.

\paragraph*{Ablation study}
To demonstrate the effectiveness in retrieving relevant spare parts based on the textual input of the user, the ranking of the proposed spare parts by the search engine. Ordering of relevant spare parts output by the search engine is evaluated in terms of top-$k$, i.e., the capability of including relevant spare parts among top-$k$ results in the ranking. The evaluation has been done for several values of $k = 1, 5, 10, 15, 20$. To better understand and quantify the contribution of each component to the system, the evaluation involves the tests of all three kind of retrievals, namely, lexical, semantic, hybrid, and the whole PhRAG system, in a supervised way employing the VSPool dataset. 
Results are presented in Tab.~\ref{tab:abl_study}.
\begin{table}[htbp]
\centering
\resizebox{\linewidth}{!}{
\begin{tabular}{lccccc}
    \toprule
    & \multicolumn{5}{c}{Components}\\ \cmidrule(lr){2-6} 
    & \multicolumn{4}{c}{Retrieval only}\\ \cmidrule(lr){2-5} 
    & \makecell{Lexical \\ (reference only)} & Semantic & \makecell{Lexical \\ (benchmark)}  & Hybrid & \makecell{PhRAG\\(hybrid \& reranker)} \\
    \midrule
    A@1  & $0.167$ & $0.693$ & $0.736$   & $0.78$ &   \underline{$0.805$} \\
    A@5  & $0.188$ & $0.799$ & $0.824$   & $0.838$ & \underline{$0.857$} \\
    A@10  & $0.193$& $0.827$  & $0.845$  & $0.853$ & \underline{$0.866$} \\
    A@15  & $0.203$& $0.834$  & $0.851$  & $0.855$ &  \underline{$0.868$} \\
    A@20  & $0.209$& $0.839$  & $0.852$  & $0.855$ & \underline{$0.869$} \\
    \bottomrule
\end{tabular}
}
\caption{Ablation study:\\ spare parts found within top-$k$ results ($k=1,5,10,15,20$)}
\label{tab:abl_study}
\end{table}

The several components of the search engine are detailed in Sec.~\ref{subsec:sea_method}.
The lexical search approach based exclusively on product reference ids effectively deals with complete, partial, or partially incorrect references of spare parts, but less than $21\%$ of the cases include the correct part among the top-$20$ positions of the provided rankings.
The standalone semantic approach is promising; it captures contextual information regarding the needs of the operator but loses details regarding references and specific technicalities. 
Extending the lexical retrieval to product descriptions and technical specifications, the search engine ensures to rank the correct components at the first position in around $73\%$ of the cases, and more than $84\%$ in the top-ten ranked spare parts. 
The hybrid search encapsulates, via ranking fusion, the set of spare parts given by the lexical and the semantic search.
Reranking via a generative language model not only improves performance but also provides an explicit justification of the retrieved components by reviewing the ranked spare parts against the query and the operator’s requirements. This latter capability, although not directly quantified, represents a significant advance over the state of the art, as traditional systems typically lack mechanisms to provide transparent reasoning or explanatory validation for the retrieved results.

From an overall view of the architecture, extraction errors may result in incomplete or incorrect structured fields within VSPool records. The search engine mitigates this risk including in the retrieval index unstructured free-text fields (name and description) independent of NER output, and reranking via the language-model based on the natural language content rather than structured fields alone. An evaluation of how extraction errors propagate into retrieval performance, for example, by injecting controlled annotation noise into VSPool and measuring the resulting retrieval degradation, will be a direction for future work.
\section{CONCLUSION}\label{sec:conclusion}
This paper introduces PhRAG, a hybrid RAG framework designed to leverage the generative multitasking capabilities of language models for technical information extraction and component retrieval in the manufacturing domain. It integrates generative models with hybrid retrieval mechanisms to support two complementary tasks: ($i$) the construction of a virtual spare-parts pool that consolidates heterogeneous unstructured textual data into a unified representation, and ($ii$) the implementation of a search engine for component retrieval from unstructured user queries. The framework, evaluated on benchmark datasets, shows improvements in performance across models of different sizes. 
The results demonstrate that PhRAG effectively addresses both objectives by employing shared components within a modular framework, without the need for fine-tuning. 
The method boosts base model performance up to fine-tuned approach capabilities in scenarios with limited annotated data and illustrates the value of generative models in refining search engine results while also providing justifications of the retrieved components for the user.






\section*{ACKNOWLEDGMENT}
This research has been partially funded by the EIT manufacturing project VSPOOL: Virtual Stock Pooling Platform (grant ID: 24292).


\printbibliography

\end{document}